%% file: ms.tex
\pdfoutput=1
\documentclass[runningheads]{llncs}

\usepackage{microtype}
\usepackage{times}
\usepackage{caption}
\usepackage{graphicx}
\usepackage{subcaption}
\usepackage{amsmath}
\usepackage{hyperref}
\usepackage{tabularx, booktabs}
\usepackage{dblfloatfix}
\usepackage{microtype}
\usepackage{algorithmic}
\usepackage{textcomp}
\usepackage{xcolor}
\usepackage{gensymb}
\usepackage{mwe}
\usepackage{amsfonts}
\usepackage{amssymb}
\usepackage{verbatim}
\usepackage{amsbsy}
\usepackage{siunitx}
\usepackage{cite}

\begin{document}

\title{Bibliographic Analysis of the Capacity and Applicability of Li-Fi Networks}
\input{0_data.tex}

\maketitle              % typeset the header of the contribution

\input{1_abstract.tex}
\input{2_introduction.tex}
\input{3_related_works.tex}
\input{4_theorethical.tex}
\input{5_methodology.tex}
\input{6_results.tex}
\input{7_conclusion.tex}
\input{references.tex}

\end{document}

%% file: 0_data.tex
\author{Kelvin I. Seibt\inst{1}\orcidID{0000-0002-0918-5983} \and
Victor A. Kich\inst{2}\orcidID{0000-0002-0547-5510} \and
Gabriel V. Heisler\inst{2}\orcidID{0000-0003-4796-9989}}

\authorrunning{Seibt. Kelvin I. et al.}

\institute{University of Santa Cruz do Sul, Santa Cruz, Brazil \\
\email{kelvin.seibt@gmail.com}\\
\and
Federal University of Santa Maria, Santa Maria, Brazil \\
\email{victorkich@yahoo.com.br} and \email{gabriel.heisler1@gmail.com}}

%% file: 1_abstract.tex
\begin{abstract}
\label{sec:abstract}
This article aims to bring a metric of capacities and applicability of Light Fidelity (Li-Fi) as a communication system through a direct comparison to subareas of Wireless Fidelity (Wi-Fi) communication, that, actually, is the main wireless communication system. The actual vision of Li-fi in a global scenario was discussed based on the use cases of this technology and your relation to 5th Generation (5G) technology. Finally, the difficulties and new paths that this system must still take were presented.\par

\end{abstract}

%% file: 2_introduction.tex
\section{Introduction}
\label{sec:introduction}
With the increase in the number of devices based on wireless communication due to the growth of the IoT (Internet of Things) area, the amount of consumed bandwidth is increasing in the same proportion. The most affected environments are public places where are a huge number of users and maybe a lot of obstacles, but depending on the case even our own home will have limitations that can be imposts on the Wi-Fi network, like interference, attenuation. Such things can harm the user experience \cite{khandal:14}.

These problems arise from many fonts, but one of the principal factors that demand attention is the limit of the spectrum of the radio waves. Realizing that the spectrum of radio waves is saturated, and for consequence is expensive, we begin to think that technologies can come to solve or minimize these imposed limits. So, our research before initiated, was a concern to find an environment with easy implementation and promissory results. In this way, was found the Li-Fi, an optic system of data communication.

Considering that all the workplaces require by law an appropriate illumination and that the electricity is a basic service in all world's countries, the Li-Fi shows to be suitable to mitigate these problems. The Li-Fi is a system that sends data through the visible and infrared (IR) light spectrum, and according to the first public demonstration of this technology, performed by professor Harald Haas in the year 2011 in his lecture at TED Talks, an environment with a light-emitting diode (LED) lamp could provide a $10$ Mbps transfer rate \cite{haas:11}.

Having found this technology, this research has as objective to evaluate if the Li-Fi could act as a substitute for Wi-Fi or only as a complementary technology. It Will be discussed areas that already have any applications to the technology, and the verdict will be made by listing comparisons with the Wi-Fi and exploring the deficiencies of this technology.

This evaluation will be done having as basis articles selected during the time frame from July 2011 to January 2020, considering the information of the book released in 2015 by the professor Haas collaborative with Svilen Dimitrov, called ``Principles of LED Light Communications: Towards Networked Li-Fi" \cite{haas:15}.

%% file: 3_related_works.tex
\section{Related Works}
\label{sec:related_works}

Three research bases were used for this search to quantify the commitment in this area that we are addressing. Were used four terms in the searches, being them: Li-Fi, Wi-Fi, 5G, and efficiency. Was adopted as a filter to the works the type ``Article" and having your publication date frame from  July of  2011  to  January of  2020 and who contains in your content the terms cited above. The bases that were used was Coordenação de Aperfeiçoamento de Pessoal de Nível Superior (CAPES), Google Scholar, and the Institute of Electrical and Electronics Engineers (IEEE).

\begin{table}[t]
\centering
\caption{Bibliometry applied to CAPES periodicals repository.}
\label{tab:exTable1}
\begin{tabular}{l|c c c c c} 
\toprule
\textbf{Term} & \textbf{Li-Fi} & \textbf{Wi-Fi} & \textbf{5G} & \textbf{Performance} \\
\midrule
\textbf{Li-Fi} & 1.986 & 85 & 65 & 18 \\
\textbf{Wi-Fi} & - & 60.026 & 1.413 & 754 \\
\textbf{5G} & - & - & 109.111 & 24.152 \\
\textbf{Performance} & - & - & - & 3.607.200 \\
\bottomrule
\end{tabular}
\end{table}

On a CAPES basis, when the term ``Li-Fi" was searched individually generated $1.986$ results. By adding the term ``Wi-Fi" to the search to get more works that would make a comparison between the technologies, this number has been reduced to $85$. Then ``5G" was added to have a return of more recent articles that were already making a connection with the future that has not yet been experienced, resulting in $65$ results. Finally, the term ``Efficiency" was added in an attempt to locate research that would quantify the results in a real environment. This is showed in Table \ref{tab:exTable1}.

The previous process was repeated in the Google Scholar and IEEE bases, presenting more promising results on Google Scholar, where it was assumed that the reason would be the broader search engine. See this in Table \ref{tab:exTable2} and \ref{tab:exTable3}.

\begin{table}[ht]
\centering
\caption{Bibliometrics applied to Google Scholar search tool.}
\label{tab:exTable2}
\begin{tabular}{l|c c c c c} 
\toprule
\textbf{Term} & \textbf{Li-Fi} & \textbf{Wi-Fi} & \textbf{5G} & \textbf{Performance} \\
\midrule
\textbf{Li-Fi} & 50.200 & 3.010 & 814 & 512 \\
\textbf{Wi-Fi} & - & 880.000 & 38.300 & 25.600 \\
\textbf{5G} & - & - & 1.450.000 & 397.000 \\
\textbf{Performance} & - & - & - & 6.000.000 \\
\bottomrule
\end{tabular}
\end{table}

\begin{table}[ht]
\centering
\caption{Bibliometry applied to IEEE Explorer research base.}
\label{tab:exTable3}
\begin{tabular}{l|c c c c c} 
\toprule
\textbf{Term} & \textbf{Li-Fi} & \textbf{Wi-Fi} & \textbf{5G} & \textbf{Performance} \\
\midrule
\textbf{Li-Fi} & 262 & 42 & 8 & 0 \\
\textbf{Wi-Fi} & - & 53 & 2 & 0 \\
\textbf{5G} & - & - & 367 & 91 \\
\textbf{Performance} & - & - & - & 3.147 \\
\bottomrule
\end{tabular}
\end{table}

As one perceives a very great influence of the creator of the technology, Haas, we find ourselves in the permission to search in other article bases, specifically by this author. Used also was his book released in 2015, called ``Principles of LED Light Communications: Towards Networked Li-Fi" \cite{haas:15}. Therefore, the ScienceDirect research base has been added to our data set. In the following paragraphs, the articles and other references used are listed.

The article proposed by Khandal \cite{khandal:14} is an analysis of the state of the technology before the launch of professor Haas' book and follows a similar path to the one approached in this research. As registered in the year 2014, offers a limited view of the progress made in this area, so it intended to update the information contained therein, using Haas' article \cite{haas:18} and his book \cite{haas:15}. After it, Bao et al. \cite{bao2015li} solve the key technologies for realizing Li-Fi and present the state-of-the-art on each aspect. Posteriorly, Wu et al. \cite{wu2017access} discussed the differences between homogeneous and heterogeneous networks regarding access point selection (APS), beyond proposing a two-stage APS method for hybrid Li-Fi/Wi-Fi networks.

The book was written by Haas and Dimitrov \cite{haas:15}, has played a key role in the construction of research since it has allowed building fundamental concepts for the initial understanding of the resources that are used in this technology. It also provided information on the evolution of the technology, since it can be compared with the article released by Professor Haas in 2018.

The article was published by professor Haas in 2018 \cite{haas:18}, where he addresses a more current view than the first article previously cited and also introduced why Li-Fi is considered a fifth-generation technology. It is intended to compare the results with those obtained by Khandal and Jain in 2014. Subsequently, Islim et al. \cite{islim2019modulation} solve the suitable modulation techniques for Li-Fi including those which explore time, frequency, and color domains.

Finally, the last selected article was A Review Paper on Li-Fi Technology \cite{singh:20}. This article was produced in the year 2020, the year in which this research is being conducted, and therefore brings us analysis in the latest updates in this technology, therefore intend to analyze it against the work of Haas \cite{haas:18}.

Given the above information in sets with the tabulated data, it has been concluded that the subject of this research has not yet been widely explored and, therefore, there is still a wide range of research that can be applied in the area.

%% file: 4_theorethical.tex
\section{Theorethical Background}
\label{sec:theorethicalbackground}

To give the reader some background, we wanted to introduce some terms and mechanisms since these will be vital for the understanding of the discussions held throughout the work.

The sources of information about Li-Fi's communication and technology came from Haas' book \cite{haas:15} and conference \cite{haas:18}. The photodiode was presented with the help of data contained in the Hamamatsu datasheet \cite{ham:S6801} and the contents of the UFRGS classes \cite{ufrgs:13}.

\subsection{Optical Communication}
\label{sub:optical}

Optical communication is any form of telecommunication that uses light as the means of transmission. Originally called Optical Wireless Communication (OWC), it has evolved as a high-capacity technology, complementing the radio frequency communications. OWC systems use wavelength in the IR spectrum for IR communications and visible light spectrum in Visible Light Communication (VLC) \cite{haas:15}.

\subsection{Photodiode}
\label{sub:photodiode}

Consists of an electronic device made of a semiconductor material (usually silicon). Have a semiconductor junction, which has the property of varying its electrical resistance according to the intensity of the light (number of photons) in its incident \cite{ufrgs:13}. The Hamamatsu Photonics company is a producer of this type of component and has models with specific applications and properties for various types of environment \cite{ham:S6801}.

\subsection{Li-Fi}
\label{sub:lifi}

Li-Fi is a VLC technology developed in 2010 through research conducted at the Edinburgh University and led by professor Haas of Mobile Communications. Became public the concept in 2011, when Haas presented the technology in his TED Talks, with the title ``Wireless data from every light bulb” \cite{haas:11}.

In 2011 the IEEE released a standard for VLC, called IEEE 802.15.7-2011, ``IEEE Standard for Local and Metropolitan Area Networks, Part 15.7: Short-Range Wireless Optical Communication Using Visible Light” \cite{haas:15}. In Li-Fi, the data communication occurs basically by changing the light intensity of a LED, which in turn is modulated into a message signal and transmitted by the wireless optical channel and detected by a photodiode, which makes the demodulation and the recovery of the message clock \cite{haas:15}.

%% file: 5_methodology.tex
\section{Methodology}
\label{sec:methodology}

This study was applied through scientific research on a worldwide basis, applying a bibliometry according to the terms that were considered central to our theme. Thus, a foundation will be made in ideas and assumptions taken from conferences, articles, and books that present importance in the construction of the concepts of this work.

The observation method used was the conceptual-analytic, since concepts and ideas from other authors will be used, which are similar to our objectives, adding new content to them for the construction of scientific analysis on the object of study.

The comparison and discussion of the founded results will be carried out by means of explanatory research, giving more freedom to talk about an analysis that traverses several communication properties and allows them to assume more than one position on the subject during the analysis.

%% file: 6_results.tex
\section{Results}
\label{sec:results}

According to Haas \cite{haas:15} the central idea of a Li-Fi wireless network is to complement heterogeneous wireless networks with radio, thus performing a relief for this spectrum and its amount of data traffic. In this way, the author already refers us to a partial response to the objective of this research, since it induces the reader that this technology will always need support, serving only as an ``end" of the network.

In the following subtopics, we wanted to have a more specific view of what the limitations may make the Li-Fi a support tool for the network and not a complete structure.

\subsection{Signal Modulation}
\label{sub:signal_modulation}

A fully optical wireless network would require omnipresent coverage by optical peripherals. The likely candidates as peripherals in VLC would be incoherent light LEDs because of their low cost. Due to the physical capacity of the components used, the information can only be encoded using light intensity. As a result, VLC can be used as an Intensity Modulation and Direct Detection (IM/DD) system, and this technique allows communication even if the lights appear visually switched off. The problem encountered with this modulation is that the signal must be real, unipolar, and not negative.

These premises eventually limit the consolidated modulation schemes used in radio wave communication. Techniques such as Pulse-Width Modulation (PWM), Pulse-Position Modulation (PPM), On-Off Keying (OOK), and Pulse-amplitude modulation (PAM) can be applied almost normally but, a modulation speeds increase, these schemes will suffer Inter-Symbol Interference (ISI) effects due to data frequency must be collected on the optical wireless network.

So a more resilient technique like Orthogonal Frequency-Division Multiplexing (OFDM) is needed, as it allows the use of adaptive bit and energy loading of different frequency sub-bands according to the properties of the communication channel. Besides, OFDMA is used in the 4th generation (4G) Long Term Evolution (LTE) communication standard for mobile phones \cite{haas:15}. Therefore, the application of OFDM in optical mobile networks would allow the use of the already established higher-level communication protocols used in IEEE 802.11 and LTE \cite{haas:15}.

At this point, both the book and Haas' article remained with the same problem, but in 2018 two solutions are already proposed: enhanced unipolar OFDM (eU-OFDM) and Spectral and Energy Efficient OFDM (SEE OFDM). There were no relevant or divergent data reported on modulation in the articles by Khandal \cite{khandal:14} and Jain and Singh \cite{singh:20}.

\subsection{Multi-user Access}
\label{sub:multiuser_access}

A fully optical, seamless network solution can only be achieved with a suitable multiple access scheme that allows multiple users to share communication resources without any mutual cross-talk. The various access schemes used in RF communications can be adapted to the OWC, provided the necessary IM/DD related modifications are made.

OFDM comes with a natural extension for multiple access OFDMA. Single-carrier modulation schemes, such as PPM and PAM, require an additional multiple access technique, such as Frequency Division Multiple Access (FDMA), Time Division Multiple Access (TDMA), or Code Division Multiple Access (CDMA) \cite{haas:15}.

In the OWC, there is an alternative option to achieve multiple access, which is the color of the LED, and the corresponding technique for this is Wavelength Division Multiple Access (WDMA). This scheme can reduce the complexity of signal processing at the expense of increasing hardware complexity.

In this sense, no changes were reported in any of the articles, remaining in this format until the present moment. One of the factors that may harm the experience of multiple users on a Li-Fi topology will be treated in subsection \ref{sub:cells}.

\subsection{Network Structure}
\label{sub:network_structure}

Ready-to-use technologies such as Power Line Communication (PLC) and Power over Ethernet (PoE) are viable back-haul solutions for Li-Fi installations \cite{haas:15}. In this sense, Wi-Fi ends up having a greater resilience, because it does not depend on a structure that uses data and energy in its communication media.

Li-Fi then ends up being limited to a range of structures so that it can operate in such a way that there is no restructuring of the network at the site. No changes were reported in any of the articles, remaining in this format until the present moment.

\subsection{Upload}
\label{sub:upload}

So far, research has focused mainly on maximizing transmission speeds on a single unidirectional link. However, for a complete Li-Fi communication system, full-duplex communication is required, i.e. an up-link connection from the mobile terminals to the optical Access Point (AP) must be provided. Existing duplex techniques used in RF such as Time Division Duplexing (TDD) and Frequency Division Duplexing (FDD) can be considered, where the downlink and up-link are separated by different time intervals or frequency ranges, respectively. However, the FDD is more difficult to perceive due to the limited bandwidth of front-end devices and because the super hetero-dynamic is not used in IM/DD systems.

The most appropriate duplexing technique in Li-Fi is Wavelength Division Duplexing (WDD), where the two communication channels are established by different electromagnetic wavelengths or using IR transmission as a viable option to establish an uplink communication channel \cite{haas:15}.

A first commercially available full-duplex Li-Fi modem using IR light for the up-link channel was recently announced by pureLiFi \cite{pure:213}. The author still reminds us that a router that needed a visible light to send the data would also be something that would generate distraction \cite{haas:18}.

This becomes relevant as there is an imbalance in traffic in favor of down-link in current wireless communication systems.

\subsection{Topology}
\label{sub:topology}

In Li-Fi the topology adopted is point-to-point, while Wi-Fi is multi-point \cite{singh:20}.

While Singh \cite{singh:20} says the technology needs a line of sight, Haas \cite{haas:15} complements that a point-to-point redirection can be done to make communication possible in a certain way, not targeted. Techniques for a really non-target vision are still being studied and also depend on the surface of reflection.

\subsection{Cells}
\label{sub:cells}

In the past, wireless cellular communication has benefited significantly from reducing the distance between stations of cellular base stations. By reducing cell size, network spectral efficiency has increased by two orders of magnitude over the past 25 years. 

More recently, different cell layers composed of microcells, picocells, and femtocells have been introduced. These networks are called heterogeneous networks. Femtocells are short-range base stations, low transmission power, low cost, and plug-and-play that are targeted for internal deployment to improve coverage. They use cable Internet or Digital Subscriber Broadband (DSL) to return to the operator's main network. The deployment of femtocells increases frequency reuse and, therefore, the transfer rate per unit area within the system, as they usually share the same bandwidth with the macrocellular network. 

However, uncoordinated and random deployment of small cells also causes additional inter and intracell interference, which imposes a limit on how dense these small base stations can be deployed before the interference begins to compensate for all the frequency reuse gains. 

The concept of small cells, however, can easily be extended to VLC to overcome the high interference generated by near-spectrum radiofrequency reuse in heterogeneous networks. Optical AP is referred to as an attoc cell. Since it operates in the optical spectrum, the optical cell does not interfere with the macrocellular network. The optic cell not only improves the internal coverage but as it does not generate any additional interference, it is capable of enhancing the capacity of radio frequency wireless networks \cite{haas:15}.

The range of cells in the 5G of technology will probably be up to $25$ meters \cite{haas:18}. This is information that comes according to the idea of Haas in 2015 when he said that the intention was that these Li-Fi cells have a maximum range of $5$ meters, so there is no overlapping of lights.

\subsection{Data Rates}
\label{sub:date_rates}

The data rate achieved by technology depends directly on the optical technology employed. Currently, most commercialized LEDs are composed of a high brightness blue LED with a phosphor cover, which converts this blue light into a yellow tone. When these two colors combine, they form the white light. This is the most cost-effective way to produce the equipment, but this conversion, which decreases the frequency of response and high frequencies, are greatly attenuated. Consequently, the bandwidth of this type of LED is around $2$ Mhz \cite{haas:18}.

However, by ignoring the cost and looking for a higher shipping fee, we can achieve incredible values. By following a speed scale, adding a blue filter to the receiver in the scheme reported in the previous paragraph it is possible to reach metrics of $1$ Gbps.

\begin{figure}[ht]
\includegraphics[width=\columnwidth]{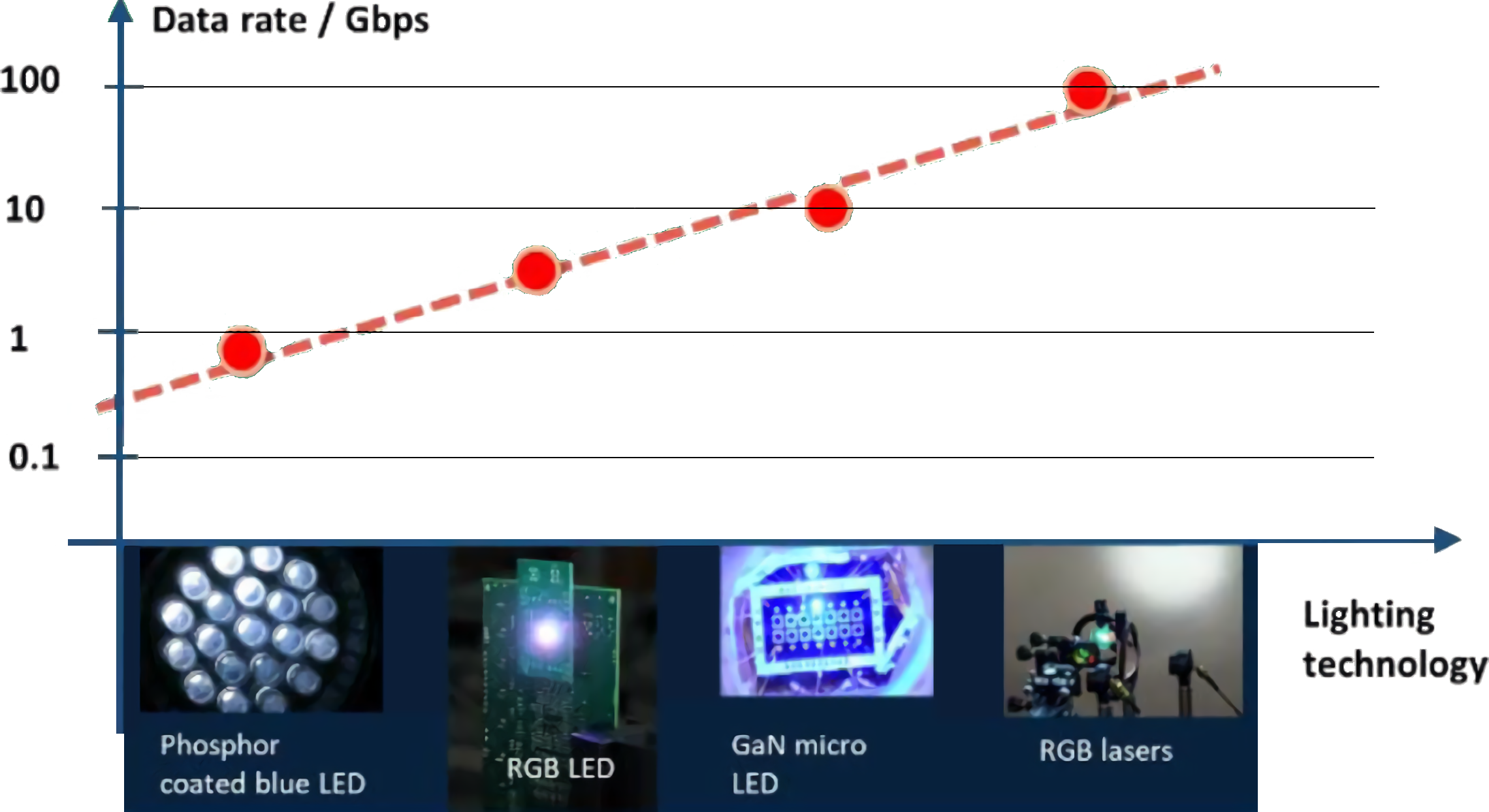}
\caption{Graph of the technologies that can be employed with VLC and its data rates, from \cite{haas:18}.}
\label{fig:Figura1}
\end{figure}

With red, green and blue (RGB) LEDs up to $8$ Gbps is achieved, as the light is naturally generated through the mixing of colors and not through chemical components. Finally, the most promising structure, laser-based lights, which have shown that a range of $100$ Gbps is achievable according to \cite{haas:18}. This is showed in Fig.\ref{fig:Figura1}.

\subsection{Use Cases}
\label{sub:use_cases}

The cases of use presented by Khandal and Jain \cite{khandal:14}, as well as Haas in his book \cite{haas:15}, are closed office or home environments, i.e. short distances. However, in his latest article \cite{haas:18} he already brings us a look, focused also on safety, pointing out the use in intrinsically safe environments, such as petrochemical plants and oil platforms where radio frequency is often forbidden. Singh finally brings us applications in the underwater environment, which is convenient with the idea of platforms in the middle of the sea, from Haas \cite{haas:18}. The light can then pass through the saltwater and works in dense regions as a sensor \cite{singh:20}.

\subsection{Li-Fi and 5G}
\label{sub:lifi_5g}

\begin{figure}[htbp]
\centerline{\includegraphics[width=\columnwidth]{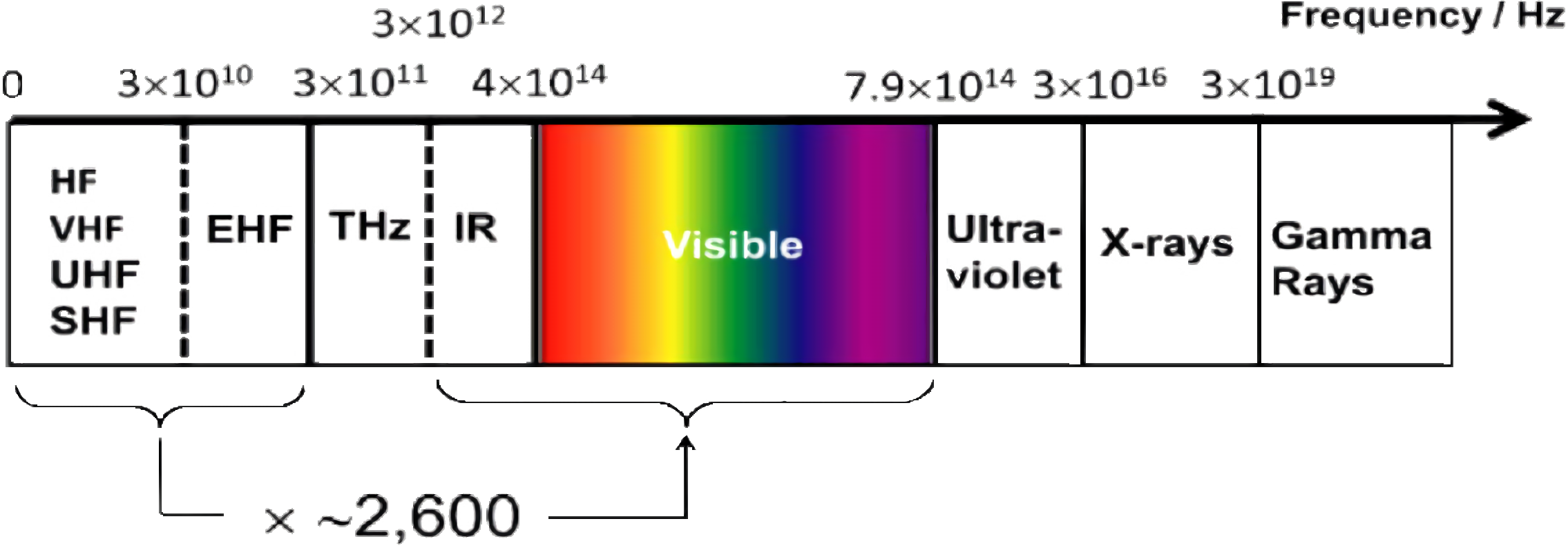}}
\caption{The RF spectrum is only a fraction of the entire electromagnetic spectrum, from \cite{haas:18}. The visible light spectrum and the IR spectrum are unregulated and offer $780$ THz bandwidth.}
\label{fig:Figura2}
\end{figure}

Currently, most technologies use RF waves for data transmission, but the growing demand for data traffic requires increasing bandwidth, which is not supported by RF. The visible light and IR specter are approximately $2.600$ times the size of the entire $300$ GHz radio frequency spectrum, appearing as a viable alternative to meet the demand for data traffic. This is possible to see in Fig.\ref{fig:Figura2}.

According to Haas \cite{haas:18} Li-Fi is a technology that will impact many industries. With the high data rate that was arranged in the subtopic ``Data Rate" and the bandwidth discussed in the paragraph above, Li-Fi fits as a 5G technology. It also adds that the adoption of the OWC standard, together with the use of RGB lasers can bring a new revolution to the area of communication. 

There would then be a large-scale unlocking of IoT in industry 4.0 applications, as well as the growth of projects that are currently just being born. An example of Light as a Service (LaaS) in the power industry, merging two major industries: the wireless communication industry and the power industry. In the near future, the LED light lamp will serve thousands of applications and will also be an integral part of smart cities, autonomous cars, smart homes, and IoT.

However, this technology is currently under development and needs more resilience in other environments where a quality connection is required, in addition to overcoming the barriers imposed by the optical medium itself and the technology employed.

%% file: 7_conclusion.tex
\section{Conclusion}
\label{sec:conclusion}

As Haas approached at the beginning of his book, Li-Fi is still just a Wi-Fi supporting technology and still has a long way to go to replace it. The complexity added to hardware to accomplish a good multi-user access and data rates make it a barrier to entry for ordinary consumers.

The point-to-point topology points that multiply spaces in the same build will need a previous preparation of network scheme. Without that, it becomes necessary for the acquisition of multiply devices to make possible the redirect of the signal. This brings up high use of hardware that, with the necessity of evolution of methodologies applied to Li-fi, can quickly become obsolete, bringing an accumulation of electronic waste.

However, it can already be observed that it is, in fact, a 5G technology with high transmission and bandwidth rates.

Based on the selected articles, the technology has not yet undergone many changes since the regulation and creation of a protocol for VLC communication. The same problems are still perceived, but some research results to add more solutions to these problems are already visible. 

Based on the results and tests cited by Haas, there is a great bet that the market will receive more prototypes and companies focused on this market, such as pureLiFi. The existing technology today is still very affected by the interference between its communication peripherals and the noise by the particularity of sunlight impulses.

With the consolidation of this technology, you will have the possibility of using in places that are impossible to use Wi-Fi. Safety is a beneficial side, where the use of light to transmit data because it does not penetrate through walls. On highways for traffic control applications, such as cars, they can have LED lights and can communicate with each other and predict accidents.

Therefore, the key point for the search for solutions to problems in Li-Fi and your implementation seems to forward automatically with the passage of time, the lack of radio band. Haas already made alarming predictions of increased data use in 2015, and today it is a reality. We have some devices that demand communication with other devices in the environment, make queries in the cloud, and still, have advanced artificial intelligence algorithms. This is a perfect environment for Li-Fi to act and its potential to be explored.

After reading the book, it is noted that there are still a large number of topics to be addressed within this technology. What stood out in terms of the need for study are the areas of signal modulation and access to multi-users. Therefore, themes directed to the discussion or solution of these deficiencies would have much relevance in the area of communication.

%% file: references.tex
\bibliographystyle{splncs04}
\bibliography{bibliography}